\documentclass[pra,twocolumn,showpacs]{revtex4}%
\usepackage{amsfonts}
\usepackage{amsmath}
\usepackage{amssymb}
\usepackage{graphicx}%
\providecommand{\U}[1]{\protect\rule{.1in}{.1in}}
\providecommand{\U}[1]{\protect\rule{.1in}{.1in}}
\providecommand{\U}[1]{\protect\rule{.1in}{.1in}}

\begin{document}
\title{Decoherence under many-body system-environment interactions: a stroboscopic
representation based on a fictitiously homogenized interaction rate}
\author{Gonzalo A. \'{A}lvarez}
\affiliation{Facultad de Matem\'{a}tica, Astronom\'{\i}a y F\'{\i}sica, Universidad
Nacional de C\'{o}rdoba, 5000 C\'{o}rdoba, Argentina}
\author{Ernesto P. Danieli}
\affiliation{Facultad de Matem\'{a}tica, Astronom\'{\i}a y F\'{\i}sica, Universidad
Nacional de C\'{o}rdoba, 5000 C\'{o}rdoba, Argentina}
\author{Patricia R. Levstein}
\affiliation{Facultad de Matem\'{a}tica, Astronom\'{\i}a y F\'{\i}sica, Universidad
Nacional de C\'{o}rdoba, 5000 C\'{o}rdoba, Argentina}
\author{Horacio M. Pastawski}
\email{horacio@famaf.unc.edu.ar}
\affiliation{Facultad de Matem\'{a}tica, Astronom\'{\i}a y F\'{\i}sica, Universidad
Nacional de C\'{o}rdoba, 5000 C\'{o}rdoba, Argentina}

\begin{abstract}
An environment interacting with portions of a system leads to multiexponential
interaction rates. Within the Keldysh formalism, we fictitiously homogenize
the system-environment interaction yielding a uniform decay rate facilitating
the evaluation of the propagators. Through an injection procedure we
neutralize the fictitious interactions. This technique justifies a
stroboscopic representation of the system-environment interaction which is
useful for numerical implementation and converges to the natural continuous
process. We apply this procedure to a fermionic two-level system and use the
Jordan-Wigner transformation to solve a two-spin swapping gate in the presence
of a spin environment.

\end{abstract}
\maketitle

\section{Introduction}

The control of open quantum systems has a fundamental relevance for fields
ranging from quantum information processing (QIP) \cite{BD2000} to
nanotechnology \cite{Awschalom2003,Taylor2003,Petta2005}. Typically, the
system whose coherent dynamics one wants to manipulate, interacts with an
environment that smoothly degrades its quantum dynamics. This process,\ called
\textquotedblleft decoherence\textquotedblright, can even be assisted by the
own system's complexity \cite{PhysicaA}. Since environment induced decoherence
\cite{Myatt2000,Gurvitz2003,Zurek2003} constitutes the main obstacle towards
QIP, a precise understanding of its inner mechanisms
\cite{Petta2005,Taylor05,Ardavan06} is critical to develop strategies to
control the quantum dynamics.

The usual way to obtain the dynamics is to solve a generalized Liouville-von
Neumann differential equation for the reduced density matrix. There the
degrees of freedom of the environment are traced out to yield a quantum master
equation (QME) \cite{QME}. A less known alternative is provided by the Keldysh
formalism \cite{Keldysh} in the integral representation proposed by
Danielewicz \cite{Danielewicz}. On one side, it uses the well known
perturbation to infinite order in selected terms provided by the Feynman
diagrams. On the other, this integral representation has the advantage of
being able to profit from a Wigner representation for the energy-time domain.
This last representation is particularly meaningful in the fermionic case
since it allows ones to define energy states and their occupations
simultaneously with the physical time \cite{GLBE2}. In that case, one can
transform the Danielewicz equation into the generalized Landauer-B\"{u}ttiker
equation (GLBE) \cite{GLBE1,GLBE2} to solve the quantum dynamics of the
system. When the system-environment (SE) interaction is spatially homogeneous,
i.e., it has an equal interaction with each component of the system, the
dynamics becomes particularly simple because there is a uniform SE interaction
rate. However, there are many situations where one should incorporate
multiple\ rates as different subsets of the system could suffer diverse
interaction processes. While this might not possess a great challenge to the
evaluation of steady state transport properties, in quantum dynamics, one is
confronted with what appears to be a much more difficult problem. Here, we
present a procedure to convert a nonhomogeneous problem with multiple SE
interaction rates, into one that has a common rate. Through a reinjection
procedure, we instantaneously neutralize the fictitious decays restoring the
populations and, eventually, the coherences. In order to illustrate the
procedure, we apply this technique to a model that represents a single fermion
that can jump between two states while an external fermionic reservoir is
coupled to one of them. This provides decoherence due to a through space
Coulomb interaction and can feed with an extra particle through tunneling
processes. While the parameters and approximations involved in this model are
especially designed to be mapped to a problem of spin dynamics, it could also
be adapted to represent a double quantum dot charge qubit
\cite{Mucciolo-Baranger2005}.\textit{ }In that case a double dot is operated
in the gate voltage regime where there is a single electron which can jump
between the two coupled states, where only one of these states is coupled to
an electron reservoir. This inhomogeneous SE interaction yields a
multiexponential decay rate. We introduce fictitious interactions to obtain a
common interaction rate which leads to a homogeneous non-Hermitian effective
Hamiltonian. In the specific model considered, we analyze how different SE
interactions, e.g., tunneling to the leads and through space Coulomb
interaction, modify the quantum evolution. A particular advantage of the
fictitious symmetrization is that it leads naturally to a stroboscopic
representation of the SE processes. This leads to a very efficient numerical
algorithm where the quantum dynamics is obtained in a sequence of time steps.
Finally, we resort to the Jordan-Wigner mapping between fermions and spins to
apply the procedure to a spin system. This allows us to give a first-principle
derivation of the self-energies used in the stroboscopic model introduced in
Ref. \cite{JCP2006} to explain the puzzling experimental dynamics observed
\cite{JCP98} in a spin swapping gate \cite{MKBE74}.

\section{System}

\begin{figure}
[th]
\begin{center}
\includegraphics[
height=4.382in,
width=3.3486in
]%
{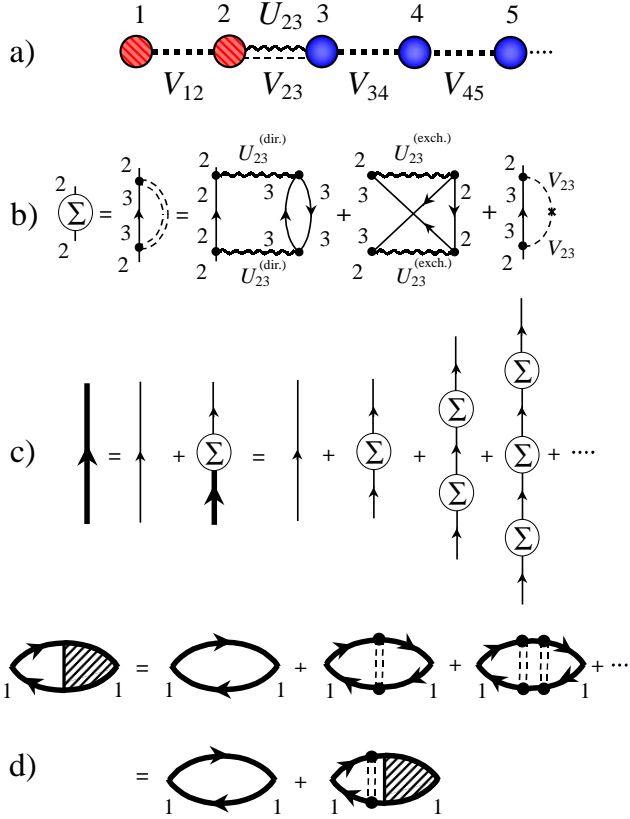}%
\caption{(Color online) (a) System-environment (SE) representation. Dashed
circles and solid circles represent the system and the environment states
respectively. Dashed lines are hopping interactions while wiggly lines are
through-space Coulomb interactions. (b) Self-energy diagram summing up the
different interactions with the environment in a local basis. The lines with
arrows are exact Green's functions in the absence of SE interactions. The
double dashed line represents the effective SE interaction. (c) Retarded
Green's function at site $1$. The interaction with the environment is to
infinite order in the self-energy given in (b). (d) Particle density function
at site $1$. The double dashed lines represent the effective interactions
local in time and space summed up to infinity order.}%
\label{Fig_system_feynman}%
\end{center}
\end{figure}

Let us consider an electron in a two-state system \ asymmetrically coupled to
an electron reservoir, as shown in Fig. \ref{Fig_system_feynman}(a), with the
total Hamiltonian $\widehat{\mathcal{H}}=\widehat{\mathcal{H}}_{\mathrm{S}%
}+\widehat{\mathcal{H}}_{\mathrm{E}}+\widehat{\mathcal{H}}_{\mathrm{SE}}%
$.\linebreak

\bigskip%

The system Hamiltonian is
\begin{equation}
\widehat{\mathcal{H}}_{\mathrm{S}}=E_{1}^{{}}\hat{c}_{1}^{+}\hat{c}_{1}^{{}%
}+E_{2}^{{}}\hat{c}_{2}^{+}\hat{c}_{2}^{{}}-V_{12}\left(  \hat{c}_{1}^{+}%
\hat{c}_{2}^{{}}+\hat{c}_{2}^{+}\hat{c}_{1}^{{}}\right)  , \label{Hs}%
\end{equation}
with $\hat{c}_{i}^{+}(\hat{c}_{i}^{{}})$ the standard fermionic creation
(destruction) operators. The $E_{i}$ are the energies of the $i$-th local
state whose spin index is omitted. The hopping interaction $V_{12}$ gives the
natural frequency $\omega_{0}=2V_{12}/\hbar$ of the transition between the
states $1$ and $2$. The environment has a similar Hamiltonian,%
\begin{equation}
\widehat{\mathcal{H}}_{\mathrm{E}}=\sum_{i=3}^{\infty}E_{i}^{{}}\hat{c}%
_{i}^{+}\hat{c}_{i}^{{}}-\sum_{%
\genfrac{}{}{0pt}{}{i,j=3}{i\neq j}%
}^{\infty}V_{ij}^{{}}\left(  \hat{c}_{i}^{+}\hat{c}_{j}^{{}}+\hat{c}_{j}%
^{+}\hat{c}_{i}^{{}}\right)  , \label{He}%
\end{equation}
where the $V_{ij}^{{}}$ determines the topology of the interaction network in
the environment states. The system-environment interaction is described by%
\begin{multline}
\widehat{\mathcal{H}}_{\mathrm{SE}}=\sum_{\alpha=\uparrow,\downarrow}\left\{
\sum_{\beta=\uparrow,\downarrow}U_{23}^{(\mathrm{dir.})}~\hat{c}_{2\beta}%
^{+}\hat{c}_{2\beta}^{{}}\hat{c}_{3\alpha}^{+}\hat{c}_{3\alpha}^{{}}\right.
\label{HSE}\\
+\left.  U_{23}^{(\mathrm{exch.})}~\hat{c}_{2\alpha}^{+}\hat{c}_{3\alpha}^{{}%
}\hat{c}_{3\alpha}^{+}\hat{c}_{2\alpha}^{{}}-V_{23}^{{}}\left(  \hat
{c}_{2\alpha}^{+}\hat{c}_{3\alpha}^{{}}+\hat{c}_{3\alpha}^{+}\hat{c}_{2\alpha
}^{{}}\right)  \right\}  ,
\end{multline}
The first two terms on the rhs represent the Coulomb interaction of an
electron in site $2$ with an electron in site $3,$ the first site of the
reservoir. $U_{23}^{(\mathrm{dir.})}$is the standard direct integral and
$U_{23}^{(\mathrm{exch.})}$ is the small exchange integral which we include
for completeness. The third term is the hopping interaction between sites $2$
and $3$.

\section{System evolution}

\subsection{Quantum dynamics in the Keldysh formalism}

We are interested in the study of the evolution of an initial local excitation
in the system. Let us consider the initial excitation with a particle on site
$2$ and a hole in site $1$ which is described by the non-equilibrium state,
\begin{equation}
\left\vert \Psi_{\mathrm{n.e.}}\right\rangle =\hat{c}_{2}^{+}\hat{c}_{1}^{{}%
}\left\vert \Psi_{\mathrm{eq.}}\right\rangle ,
\end{equation}
where $\left\vert \Psi_{\mathrm{eq.}}\right\rangle $ is the thermodynamical
many-body equilibrium state at high temperatures which is the regime of NMR
spin dynamics. In this condition $\left\vert \Psi_{\mathrm{eq.}}\right\rangle
$ is the mixture, with equal weight, of all the possible Slater determinants
\cite{spin-projection}. The evolution in a complete norm preserving solution
is described by the particle and hole density functions
\begin{equation}
G_{ij}^{<}\left(  t_{2},t_{1}\right)  =\frac{\mathrm{i}}{\hbar}\left\langle
\Psi_{\mathrm{n.e.}}\right\vert \hat{c}_{j}^{+}\left(  t_{1}\right)  \hat
{c}_{i}^{{}}\left(  t_{2}\right)  \left\vert \Psi_{\mathrm{n.e.}%
}\right\rangle
\end{equation}
and
\begin{equation}
G_{ij}^{>}\left(  t_{2},t_{1}\right)  =-\frac{\mathrm{i}}{\hbar}\left\langle
\Psi_{\mathrm{n.e.}}\right\vert \hat{c}_{i}^{{}}\left(  t_{2}\right)  \hat
{c}_{j}^{+}\left(  t_{1}\right)  \left\vert \Psi_{\mathrm{n.e.}}\right\rangle
,
\end{equation}
that describe spatial and temporal correlations. In these expressions, the
creation and destruction operators are in the Heisenberg representation.
Notice that in contrast with the equilibrium definitions of $G_{ij}%
^{\lessgtr\,}(t_{2},t_{1}),$ now they have an implicit dependence on the
initial local excitation. The probability amplitude of finding a particle in
site $i$ at time $t_{2}$ when it initially was in site $j$ at time $t_{1}$ is
described by the retarded Green's function of the whole system
\begin{align}
G_{ij}^{\mathrm{R}}\left(  t_{2},t_{1}\right)   &  =\theta\left(  t_{2}%
,t_{1}\right)  \ [G_{ij}^{>}\left(  t_{2},t_{1}\right)  -G_{ij}^{<}\left(
t_{2},t_{1}\right)  ]\nonumber\\
&  =\left[  G_{ji}^{\mathrm{A}}\left(  t_{1},t_{2}\right)  \right]  ^{\ast}.
\end{align}
The reduced density function $\mathbf{G}^{<}\left(  t,t\right)  $, where
matrix indices are restricted to $i,j\in\left\{  1,2\right\}  $, is equivalent
to the single particle $2\times2$ density matrix and $\mathbf{G}^{\mathrm{R}%
}\left(  t_{2},t_{1}\right)  $ is an effective evolution operator
\cite{matrix-representation}. If the system is isolated, the Green's function
in its energy representation is obtained by a Fourier transform (FT) with
respect to the time interval $\delta t=t_{2}-t_{1},$
\begin{equation}
\mathbf{G}^{0\mathrm{R}}\left(  \varepsilon,t\right)  =\int\mathbf{G}%
^{0\mathrm{R}}\left(  t+\tfrac{1}{2}\delta t,t-\tfrac{1}{2}\delta t\right)
\exp[\mathrm{i}\varepsilon\delta t/\hbar]\mathrm{d}\delta t,
\end{equation}
where $t=\frac{1}{2}\left(  t_{2}+t_{1}\right)  .~$In a time independent
system%
\begin{equation}
\mathbf{G}^{0\mathrm{R}}\left(  \varepsilon,t\right)  \equiv\mathbf{G}%
^{0\mathrm{R}}\left(  \varepsilon\right)  =[\varepsilon\mathbf{I}%
-\mathbf{H}_{\mathrm{S}}]^{-1}.
\end{equation}
After including SE interactions, the Green's function defines the reduced
effective Hamiltonian and the self-energy $\mathbf{\Sigma}^{\mathrm{R}%
}(\varepsilon)$ \cite{DAmato},\emph{ }%
\begin{equation}
\mathbf{H}_{\mathrm{eff.}}(\varepsilon)\equiv\varepsilon\mathbf{I}-\left[
\mathbf{G}^{\mathrm{R}}\left(  \varepsilon\right)  \right]  ^{-1}%
=\mathbf{H}_{\mathrm{S}}+\mathbf{\Sigma}^{\mathrm{R}}(\varepsilon).
\label{def-Heff}%
\end{equation}
Here, the exact perturbed dynamics is contained in the nonlinear dependence of
the self-energy $\mathbf{\Sigma}^{\mathrm{R}}$ on $\varepsilon$\emph{.} For
infinite reservoirs the evolution with $\mathbf{H}_{\mathrm{eff.}}$ is
nonunitary, hence, the Green's function has poles at the \textquotedblleft
eigenenergies\textquotedblright, $\varepsilon_{\nu}$, \ that have imaginary
components \cite{DAmato-Pastawski},
\begin{equation}
-2\operatorname{Im}\Sigma^{\mathrm{R}}\left(  \varepsilon_{\nu}\right)
/\hbar=1/\tau_{\mathrm{SE}}^{{}}=2\Gamma_{\mathrm{SE}}/\hbar.
\end{equation}
These account for the\ \textquotedblleft decay rates\textquotedblright\ into
collective SE eigenstates in agreement with a self-consistent Fermi golden
rule (FGR) \cite{SC-FGR}. Similarly, $\operatorname{Re}\Sigma^{\mathrm{R}%
}\left(  \varepsilon_{\nu}^{{}}\right)  =\operatorname{Re}\varepsilon_{\nu
}-\varepsilon_{\nu}^{0}$ represent the \textquotedblleft
shifts\textquotedblright\ of the system's eigenenergies $\varepsilon_{\nu}%
^{0}.$

The evolution of the density function for the reduced open system is described
using the Keldysh formalism \cite{Keldysh,Danielewicz}. The density function
in the Danielewicz form \cite{Danielewicz} is%
\begin{multline}
\mathbf{G}^{<}\left(  t_{2},t_{1}\right)  =\hbar^{2}\mathbf{G}^{\mathrm{R}%
}\left(  t_{2},0\right)  \mathbf{G}^{<}\left(  0,0\right)  \mathbf{G}%
^{\mathrm{A}}\left(  0,t_{1}\right) \label{Danielewicz_evol}\\
+\int_{0}^{t_{2}}\int_{0}^{t_{1}}\mathrm{d}t_{k}\mathrm{d}t_{l}\mathbf{G}%
^{\mathrm{R}}\left(  t_{2},t_{k}\right)  \mathbf{\Sigma}^{<}\left(
t_{k},t_{l}\right)  \mathbf{G}^{\mathrm{A}}\left(  t_{l},t_{1}\right)  .
\end{multline}
The first term is the \textquotedblleft coherent\textquotedblright\ evolution
while the second term contains \textquotedblleft incoherent
reinjections\textquotedblright\ through the self-energy function
$\mathbf{\Sigma}^{<}$. This compensates any leak from the coherent evolution
reflected by the imaginary part of $\Sigma^{\mathrm{R}}$ (see \cite{GLBE2}).
We remark that this expression is valid for a noncorrelated initial state
which is our case of interest. For a correlated state see Ref. \cite{Almbladh}%
. The key to solve Eq. (\ref{Danielewicz_evol}) is to build up an expression
for the particle (hole) injection and retarded self-energies $\mathbf{\Sigma
}^{<(>)}\left(  t_{1},t_{2}\right)  $ and
\begin{equation}
\mathbf{\Sigma}^{\mathrm{R}}\left(  t_{1},t_{2}\right)  =\theta\left(
t_{1},t_{2}\right)  [\mathbf{\Sigma}^{>}\left(  t_{2},t_{1}\right)
-\mathbf{\Sigma}^{<}\left(  t_{2},t_{1}\right)  ]. \label{SigmaR}%
\end{equation}
For this purpose, we use a perturbative expansion on $\widehat{\mathcal{H}%
}_{\mathrm{SE}}$ like that used in Ref. \cite{SSC2005}\emph{ }for the Coulomb
interaction and in Ref. \cite{CPL2005} for the hopping interaction.\emph{ }The
first order in the perturbation expansion is the standard Hartree-Fock energy
correction which does not contribute to $\Sigma^{<}$ because it is real\emph{.
}We focus on the second-order terms, with Feynman diagrams sketched in Fig.
\ref{Fig_system_feynman}(b).

The injection self-energy is\emph{ }%
\begin{multline}
\Sigma_{ij}^{\lessgtr}\left(  t_{k},t_{l}\right)  =\label{Sigma_Feynman}\\
\left\vert U_{23}^{{}}\right\vert ^{2}\hbar_{{}}^{2}G_{33}^{\lessgtr}\left(
t_{k},t_{l}\right)  G_{33}^{\gtrless}\left(  t_{l},t_{k}\right)
G_{22}^{\lessgtr}\left(  t_{k},t_{l}\right)  ~\delta_{i2}^{{}}\delta_{2j}^{{}%
}\\
+\left\vert V_{23}^{{}}\right\vert ^{2}G_{33}^{\lessgtr}\left(  t_{k}%
,t_{l}\right)  ~\delta_{i2}^{{}}\delta_{2j}^{{}},
\end{multline}
where $U_{23}=-2U_{23}^{(\mathrm{dir.})}+U_{23}^{(\mathrm{exch.})}$ is the net
Coulomb interaction between an electron in the system and one in the
reservoir. The direct term contributes with a fermion loop and an extra spin
summation which is represented in the $-2$ factor \cite{Danielewicz}. The
first term in Eq. (\ref{Sigma_Feynman}) corresponds to the direct and exchange
self-energy diagrams shown in the last line of Fig. \ref{Fig_system_feynman}%
(b). The first two diagrams schematize the creation of an electron-hole pair
in the environment and its later destruction. The last term in Eq.
(\ref{Sigma_Feynman}) and the last diagram of the same figure is the hopping
to site $3$ which allows the electron to perform a full exploration inside the
reservoir. To take into account the different time scales for the dynamics of
excitations in the system and in the reservoir, we use the energy-time
variables: the physical time $t_{\mathrm{i}}=\frac{1}{2}\left(  t_{k}%
+t_{l}\right)  ,$ and the domain of quantum correlations $\delta
t_{\mathrm{i}}=t_{k}-t_{l}.$ This last is related to an energy $\varepsilon$
through a FT \cite{GLBE2}. Thus, in equilibrium,%
\begin{align}
G_{33}^{<}\left(  \varepsilon,t_{\mathrm{i}}\right)   &  =\mathrm{i}2\pi
~N_{3}\left(  \varepsilon\right)  ~\mathrm{f}_{3}\left(  \varepsilon
,t_{\mathrm{i}}\right)  ,\label{G<(E,t)}\\
G_{33}^{>}\left(  \varepsilon,t_{\mathrm{i}}\right)   &  =-\mathrm{i}%
2\pi~N_{3}\left(  \varepsilon\right)  ~\left[  1-\mathrm{f}_{3}\left(
\varepsilon,t_{\mathrm{i}}\right)  \right]  ,
\end{align}
where $N_{3}\left(  \varepsilon\right)  $ is the local density of states
(LDoS)\ at the surface of the reservoir. Assuming that the environment stays
in the thermodynamical equilibrium and $k_{\mathrm{B}}T$ is much higher than
any energy scale in the bath (high temperature limit), the occupation factor
is$\ $%
\begin{equation}
\mathrm{f}_{3}\left(  \varepsilon,t_{\mathrm{i}}\right)  =\mathrm{f}_{3}.
\end{equation}
Fourier transforming on $\varepsilon$ one obtains
\begin{equation}
G_{33}^{<}\left(  t_{\mathrm{i}}+\tfrac{\delta t_{\mathrm{i}}}{2}%
,t_{\mathrm{i}}-\tfrac{\delta t_{\mathrm{i}}}{2}\right)  =\mathrm{i}%
2\pi\mathrm{~}g_{3}\left(  \delta t_{\mathrm{i}}\right)  ~\mathrm{f}_{3}%
\end{equation}
and
\begin{equation}
G_{33}^{>}\left(  t_{\mathrm{i}}+\tfrac{\delta t_{\mathrm{i}}}{2}%
,t_{\mathrm{i}}-\tfrac{\delta t_{\mathrm{i}}}{2}\right)  =-\mathrm{i}%
2\pi\mathrm{~}g_{3}\left(  \delta t_{\mathrm{i}}\right)  ~\left(
1-\mathrm{f}_{3}\right)  ,
\end{equation}
where
\begin{equation}
g_{3}\left(  \delta t_{\mathrm{i}}\right)  =\int N_{3}\left(  \varepsilon
\right)  e^{-\mathrm{i}\varepsilon\delta t_{\mathrm{i}}}\frac{\mathrm{d}%
\varepsilon}{2\pi\hbar}.
\end{equation}
Replacing in Eq. (\ref{Sigma_Feynman})\emph{ }%
\begin{multline}
\Sigma_{ij}^{\lessgtr}\left(  t_{\mathrm{i}}+\tfrac{\delta t_{\mathrm{i}}}%
{2},t_{\mathrm{i}}-\tfrac{\delta t_{\mathrm{i}}}{2}\right)
=\label{Sigma_Feynmann_2}\\
\left\vert U_{23}^{{}}\right\vert ^{2}\hbar_{{}}^{2}\left(  2\pi\right)
^{2}\left[  g_{3}\left(  \delta t_{\mathrm{i}}\right)  \right]  ^{2}%
\mathrm{f}_{3}\left[  1-\mathrm{f}_{3}\right] \\
\times G_{22}^{\lessgtr}\left(  t_{\mathrm{i}}+\tfrac{\delta t_{\mathrm{i}}%
}{2},t_{\mathrm{i}}-\tfrac{\delta t_{\mathrm{i}}}{2}\right)  \delta_{i2}^{{}%
}\delta_{2j}^{{}}\\
\pm\left\vert V_{23}^{{}}\right\vert ^{2}~\mathrm{i}2\pi g_{3}\left(  \delta
t_{\mathrm{i}}\right)  ~\left(
\genfrac{}{}{0pt}{}{\mathrm{f}_{3}}{1-\mathrm{f}_{3}}%
\right)  ~\delta_{i2}^{{}}\delta_{2j}^{{}},
\end{multline}
where the $\left(
\genfrac{}{}{0pt}{}{\mathrm{f}_{3}}{1-\mathrm{f}_{3}}%
\right)  $ associates $\mathrm{f}_{3}$ with $\Sigma^{<}$ and $\left(
1-\mathrm{f}_{3}\right)  $ with $\Sigma^{>}$.

In summary, we are left with the task to evaluate the time dependent
self-energies and the integral in Eq. (\ref{Danielewicz_evol}). We will focus
on the parametric regime corresponding to the experimental conditions of the
spin swapping gate.

\subsection{Environment in the wide band or fast fluctuations regime}

As occurs with the generalized Landauer-B\"{u}ttiker equations for linear
transport, an essential ingredient is the possibility to assign a Markovian
nature to the environment. We are going to see that this appears naturally
from the formalism when the dynamics of excitations within the environment is
faster than the time scales relevant to the system. In order to separate the
different physical time scales involved in the problem, we start changing to
the energy-time variables in Eq. (\ref{Danielewicz_evol}). Evaluating in
$t_{2}=t_{1}=t,$ the integrand becomes
\begin{multline}
\int_{0}^{t}\mathrm{d}t_{\mathrm{i}}\int_{-t}^{t}\mathrm{d}\delta
t_{\mathrm{i}}\\
\times G^{\mathrm{R}}\left(  t,t_{\mathrm{i}}+\tfrac{\delta t_{\mathrm{i}}}%
{2}\right)  \Sigma^{<}\left(  t_{\mathrm{i}}+\tfrac{\delta t_{\mathrm{i}}}%
{2},t_{\mathrm{i}}-\tfrac{\delta t_{\mathrm{i}}}{2}\right)  G^{\mathrm{A}%
}\left(  t_{\mathrm{i}}-\tfrac{\delta t_{\mathrm{i}}}{2},t\right)  .
\end{multline}
The environment unperturbed\ Green's function $g_{3}\left(  \delta
t_{\mathrm{i}}\right)  $ decays within the time scale $\hbar/V_{\mathrm{B}}$
where $V_{\mathrm{B}}$ is the characteristic interaction inside the reservoir.
In the wide band regime ($V_{\mathrm{B}}\gg V_{12}$) $\hbar/V_{\mathrm{B}}$
becomes much shorter than the characteristic evolution time of $G_{22}%
^{\lessgtr}\left(  t_{\mathrm{i}}+\tfrac{\delta t_{\mathrm{i}}}{2}%
,t_{\mathrm{i}}-\tfrac{\delta t_{\mathrm{i}}}{2}\right)  $ given by
$\hbar/V_{12}.$ Then, as the main contribution to the integral on $\delta
t_{\mathrm{i}}$ of Eq. (\ref{Danielewicz_evol}) is around the time scale
$\hbar/V_{\mathrm{B}}$ we can replace $G_{22}^{\lessgtr}\left(  t_{\mathrm{i}%
}+\tfrac{\delta t_{\mathrm{i}}}{2},t_{\mathrm{i}}-\tfrac{\delta t_{\mathrm{i}%
}}{2}\right)  $ by $G_{22}^{\lessgtr}\left(  t_{\mathrm{i}},t_{\mathrm{i}%
}\right)  $. Following the same assumption we replace $G^{\mathrm{R}}\left(
t,t_{\mathrm{i}}+\frac{\delta t_{\mathrm{i}}}{2}\right)  $ by $G^{\mathrm{R}%
}\left(  t,t_{\mathrm{i}}\right)  $ and $G^{\mathrm{A}}\left(  t_{\mathrm{i}%
}-\frac{\delta t_{\mathrm{i}}}{2},t\right)  $ by $G^{\mathrm{A}}\left(
t_{\mathrm{i}},t\right)  .$\emph{ }In this fast fluctuation regime, only
$\Sigma_{ij}^{\lessgtr}\left(  t_{\mathrm{i}}+\tfrac{\delta t_{\mathrm{i}}}%
{2},t_{\mathrm{i}}-\tfrac{\delta t_{\mathrm{i}}}{2}\right)  $ depends on
$\delta t_{\mathrm{i}}$ leading to%
\begin{multline}
\Sigma_{ij}^{\lessgtr}\left(  t_{\mathrm{i}}\right)  =\int_{-t}^{t}\Sigma
_{ij}^{\lessgtr}\left(  t_{\mathrm{i}}+\tfrac{\delta t_{\mathrm{i}}}%
{2},t_{\mathrm{i}}-\tfrac{\delta t_{\mathrm{i}}}{2}\right)  \mathrm{d}\delta
t_{\mathrm{i}}\label{Sigma_ti}\\
=\left\vert U_{23}^{{}}\right\vert ^{2}\hbar_{{}}^{2}\left(  2\pi\right)
^{2}\left[  \int_{-t}^{t}\left[  g_{3}\left(  \delta t_{\mathrm{i}}\right)
\right]  ^{2}\mathrm{d}\delta t_{\mathrm{i}}\right]  \mathrm{f}_{3}\left[
1-\mathrm{f}_{3}\right] \\
\times G_{22}^{\lessgtr}\left(  t_{\mathrm{i}},t_{\mathrm{i}}\right)
~\delta_{i2}^{{}}\delta_{2j}^{{}}\\
\pm\left\vert V_{23}^{{}}\right\vert ^{2}~\mathrm{i}2\pi\left[  \int_{-t}%
^{t}g_{3}\left(  \delta t_{\mathrm{i}}\right)  \mathrm{d}\delta t_{\mathrm{i}%
}\right]  \left(
\genfrac{}{}{0pt}{}{\mathrm{f}_{3}}{1-\mathrm{f}_{3}}%
\right)  \delta_{i2}^{{}}\delta_{2j}^{{}},
\end{multline}
which is local in space and time.\emph{ }Here, because of the limit
$V_{12}/V_{\mathrm{B}}\rightarrow0$, the correlation function of site $3$
becomes a representation of the Dirac delta function. Thus, any perturbation
at site $3$ is almost instantaneously spread all over the environment (as
compared with the time scale of the system dynamics) and hence the occupation
at site $3$ remains constant. This assumption for the time scales can be seen
in Fig. \ref{Fig_system_feynman}(b) as a collapse of a pair of black dots,
along a vertical line, into a single point. This justifies the expansion of
Fig. \ref{Fig_system_feynman}(c) and\ the use of the ladder approximation
containing only vertical interaction lines in Fig. \ref{Fig_system_feynman}(d).

A generalized decay rate is given by%
\begin{align}
1/\tau_{\mathrm{SE}}^{{}}\left(  \varepsilon,t_{\mathrm{i}}\right)   &
\equiv2\Gamma_{\mathrm{SE}}^{{}}\left(  \varepsilon,t_{\mathrm{i}}\right)
/\hbar\equiv-2\operatorname{Im}\Sigma_{{}}^{\mathrm{R}}\left(  \varepsilon
,t_{\mathrm{i}}\right)  /\hbar\label{GammaSE}\\
&  =\tfrac{\mathrm{i}}{\hbar}\left[  \Sigma_{22}^{\mathrm{A}}\left(
\varepsilon,t_{\mathrm{i}}\right)  -\Sigma_{22}^{\mathrm{R}}\left(
\varepsilon,t_{\mathrm{i}}\right)  \right] \\
&  =\tfrac{\mathrm{i}}{\hbar}\left[  \Sigma_{22}^{>}\left(  \varepsilon
,t_{\mathrm{i}}\right)  -\Sigma_{22}^{<}\left(  \varepsilon,t_{\mathrm{i}%
}\right)  \right]  ,
\end{align}
where%
\begin{equation}
\Sigma_{ij}^{\lessgtr}\left(  \varepsilon,t_{\mathrm{i}}\right)
=\int_{-\infty}^{\infty}\Sigma_{ij}^{\lessgtr}\left(  t_{\mathrm{i}}%
+\tfrac{\delta t_{\mathrm{i}}}{2},t_{\mathrm{i}}-\tfrac{\delta t_{\mathrm{i}}%
}{2}\right)  e^{\mathrm{i}\varepsilon\delta t_{\mathrm{i}}/\hbar}%
\mathrm{d}\delta t_{\mathrm{i}}. \label{Sigma_menos_mas_eti}%
\end{equation}
We start assuming $E_{i}=0$ for $i=1,..,\infty$, so the only relevant energy
scale of the system is $V_{12}\ll V_{\mathrm{B}}.$ As mentioned above, in the
wide band limit the correlation function $g_{3}\left(  \delta t_{\mathrm{i}%
}\right)  $ becomes a representation of the Dirac delta function. In this way,
the term $e^{\mathrm{i}\varepsilon\delta t_{\mathrm{i}}/\hbar}$ of the
integrand of Eq. (\ref{Sigma_menos_mas_eti}) is evaluated for $\delta
t_{\mathrm{i}}=0$ giving a value equal to $1.$ Thus, using Eq. (\ref{Sigma_ti}%
), we obtain for the decay rate which in the wide band limit is constant in
time and independent of energy,%
\begin{align}
\frac{1}{\tau_{\mathrm{SE}}}  &  \underset{\mathrm{WB}}{=}\tfrac{\mathrm{i}%
}{\hbar}[\Sigma_{22}^{>}\left(  t_{\mathrm{i}}\right)  -\Sigma_{22}^{<}\left(
t_{\mathrm{i}}\right)  ]\\
&  =\left\vert U_{23}^{{}}\right\vert ^{2}~\left(  2\pi\right)  ^{2}\left[
\int_{-t}^{t}\left[  g_{3}\left(  \delta t_{\mathrm{i}}\right)  \right]
^{2}\mathrm{d}\delta t_{\mathrm{i}}\right]  ~\mathrm{f}_{3}\left[
1-\mathrm{f}_{3}\right] \nonumber\\
&  +\tfrac{1}{\hbar}\left\vert V_{23}^{{}}\right\vert ^{2}~2\pi\left[
\int_{-t}^{t}g_{3}\left(  \delta t_{\mathrm{i}}\right)  \mathrm{d}\delta
t_{\mathrm{i}}\right] \\
&  =\tfrac{2}{\hbar}\left(  \Gamma_{U}+\Gamma_{V}\right)  ,
\label{GammaU+GammaV}%
\end{align}
where we have used $t\gg\hbar/V_{\mathrm{B}}$ to equal $\Sigma_{ij}^{\lessgtr
}\left(  \varepsilon,t_{\mathrm{i}}\right)  =\Sigma_{ij}^{\lessgtr}\left(
t_{\mathrm{i}}\right)  $ and define
\begin{equation}
\Gamma_{U}=\hbar\left\vert U_{23}^{{}}\right\vert ^{2}~2\pi^{2}\left[
\int_{-\infty}^{\infty}\left[  g_{3}\left(  \delta t_{\mathrm{i}}\right)
\right]  ^{2}\mathrm{d}\delta t_{\mathrm{i}}\right]  ~\mathrm{f}_{3}\left[
1-\mathrm{f}_{3}\right]  ,
\end{equation}
the Coulomb decay rate, and
\begin{equation}
\Gamma_{V}=\left\vert V_{23}^{{}}\right\vert ^{2}~\pi\left[  \int_{-\infty
}^{\infty}g_{3}\left(  \delta t_{\mathrm{i}}\right)  \mathrm{d}\delta
t_{\mathrm{i}}\right]  ,
\end{equation}
the hopping decay rate. If one assumes that the environment (\ref{He}) can be
represented by a linear chain with near neighbor hoppings equal to
$V_{\mathrm{B}}$ and $E_{i}\equiv0,$ the LDoS is (see Ref. \cite{CPL2005})
\begin{equation}
N_{3}\left(  \varepsilon\right)  =1/\left(  \pi V_{\mathrm{B}}\right)
\sqrt{1-\left(  \frac{\varepsilon}{2V_{\mathrm{B}}}\right)  ^{2}}.
\end{equation}
Thus, the Green's function
\begin{equation}
g_{3}\left(  \delta t_{\mathrm{i}}\right)  =\frac{1}{2\pi V_{\mathrm{B}}}%
\frac{J_{1}\left(  \frac{2V_{\mathrm{B}}}{\hbar}\delta t_{i}\right)  }{\delta
t_{i}}%
\end{equation}
is proportional to the first-order Bessel function and decays within a
characteristic time $\hbar/V_{\mathrm{B}}.$ Assuming that $\mathrm{f}_{3}=1/2$
and the integration limits in the $\Gamma$'s expressions are taken to infinity
because $t\sim\hbar/V_{12}\gg\hbar/V_{\mathrm{B}}$ (wide band approximation),
one obtains
\begin{equation}
\tfrac{2}{\hbar}\Gamma_{U}=\tfrac{2\pi}{\hbar}\left\vert U_{23}^{{}%
}\right\vert ^{2}\dfrac{2}{3\pi^{2}V_{\mathrm{B}}} \label{gammaU}%
\end{equation}
and
\begin{equation}
\tfrac{2}{\hbar}\Gamma_{V}=\tfrac{2\pi}{\hbar}\left\vert V_{23}^{{}%
}\right\vert ^{2}\frac{1}{\pi V_{\mathrm{B}}}. \label{gammaV}%
\end{equation}
Since the interaction is local in time, the reduced density results as follows%
\begin{multline}
\mathbf{G}^{<}\left(  t,t\right)  =\hbar^{2}\mathbf{G}^{\mathrm{R}}\left(
t,0\right)  \mathbf{G}^{<}\left(  0,0\right)  \mathbf{G}^{\mathrm{A}}\left(
0,t\right) \\
+\int_{0}^{t}\mathrm{d}t_{\mathrm{i}}\mathbf{G}^{\mathrm{R}}\left(
t,t_{\mathrm{i}}\right)  \mathbf{\Sigma}_{{}}^{<}\left(  t_{\mathrm{i}%
}\right)  \mathbf{G}^{\mathrm{A}}\left(  t_{\mathrm{i}},t\right)  ,
\label{GLBE-inhomogeneous}%
\end{multline}
which\ is complemented with%
\begin{equation}
\mathbf{\Sigma}_{{}}^{<}\left(  t_{\mathrm{i}}\right)  =\left(
\begin{array}
[c]{cc}%
0 & 0\\
0 & 2\Gamma_{U}^{{}}\hbar G_{22}^{<}\left(  t_{\mathrm{i}},t_{\mathrm{i}%
}\right)  +2\Gamma_{V}^{{}}\hbar\left(  \frac{\mathrm{i}}{\hbar}\mathrm{f}%
_{3}^{{}}\right)
\end{array}
\right)  .
\end{equation}
Here, the propagators $\mathbf{G}^{\mathrm{R}}\left(  t,0\right)  $ and
$\mathbf{G}^{\mathrm{A}}\left(  0,t\right)  $ that enter in both terms are
obtained from the effective Hamiltonian of the reduced system,%
\begin{equation}
\mathbf{H}_{\mathrm{eff.}}=\left(
\begin{array}
[c]{cc}%
0 & -V_{12}\\
-V_{12} & -\mathrm{i}\Gamma_{\mathrm{SE}}%
\end{array}
\right)  , \label{Heff}%
\end{equation}
where $\Gamma_{\mathrm{SE}}$ is energy independent and the assumption
$E_{i}=0$ for all $i$ assures that the self-energies are purely imaginary.
This effective Hamiltonian is obtained from Eq. (\ref{def-Heff}) and using Eq.
(\ref{SigmaR}) previously converted into the energy-time variables. This means
that we first change the variables $\left(  t_{1},t_{2}\right)  $ to $\left(
t_{\mathrm{i}},\delta t_{\mathrm{i}}\right)  $ and then based on a Fourier
transform on $\delta t_{\mathrm{i}}$ we obtain the energy-time representation
which results in the independence on both $\varepsilon$ and $t_{\mathrm{i}}$.

The above procedure results in an equation of the form of the GLBE. However,
the Hamiltonian is asymmetric in the SE interaction complicating the form of
the associated \ propagator. The apparent complexity to solve this equation
contrasts with the homogeneous case where the evolution of the GLBE was
obtained \cite{GLBE1} through a Laplace transform. Our strategy will be to
induce such a form of symmetry.

\subsection{Fictitious homogeneous decay}

The main difficulty with Eq. (\ref{GLBE-inhomogeneous}) is that it involves
multiple exponentials. In order to create propagators with an homogeneous
decay, i.e., a single exponential factor, we introduce \emph{fictitious
interactions} $\mathbf{\Sigma}_{\mathrm{fic.}}^{\mathrm{R}}$ with the
environment. The symmetric Hamiltonian becomes%
\begin{align}
\mathbf{H}_{\mathrm{sym.}}  &  =\mathbf{H}_{\mathrm{eff.}}+\mathbf{\Sigma
}_{\mathrm{fic.}}^{\mathrm{R}}\nonumber\\
&  =\left(
\begin{array}
[c]{cc}%
0 & -V_{12}\\
-V_{12} & -\mathrm{i}\Gamma_{\mathrm{SE}}%
\end{array}
\right)  +\left(
\begin{array}
[c]{cc}%
-\mathrm{i}\frac{1}{2}\Gamma_{\mathrm{SE}} & 0\\
0 & \mathrm{i}\frac{1}{2}\Gamma_{\mathrm{SE}}%
\end{array}
\right) \nonumber\\
&  =\left(
\begin{array}
[c]{cc}%
-\mathrm{i}\frac{1}{2}\Gamma_{\mathrm{SE}} & -V_{12}\\
-V_{12} & -\mathrm{i}\frac{1}{2}\Gamma_{\mathrm{SE}}%
\end{array}
\right)  . \label{Hsym}%
\end{align}
Here $\mathbf{\Sigma}_{\mathrm{fic.}}^{\mathrm{R}}$ includes the fictitious
interactions which, in the present case, produce a \textit{leak of
probability} in site $1$ at a rate $\Gamma_{\mathrm{SE}}/\hbar$ while in site
$2$ \textit{inject probability} at the same rate. Both states of
$\mathbf{H}_{\mathrm{sym.}}$ interact with the environment independently with
the same characteristic decay rate $\Gamma_{\mathrm{SE}}/\hbar.$ Note that
this rate is half the real value. The propagators of Eq.
(\ref{Danielewicz_evol}) have now a simple dependence on $t$ as
\begin{equation}
\mathbf{G}^{\mathrm{R}}\left(  t,0\right)  =\mathbf{G}^{0\mathrm{R}}\left(
t,0\right)  e^{-\frac{\Gamma_{\mathrm{SE}}}{2}t/\hbar},
\end{equation}
\emph{ }where
\begin{equation}
G_{11}^{0\mathrm{R}}(t,0)=G_{22}^{0\mathrm{R}}(t,0)=\frac{\mathrm{i}}{\hbar
}\cos\left(  \frac{\omega_{0}}{2}t\right)
\end{equation}
and
\begin{equation}
G_{12}^{0\mathrm{R}}(t,0)=G_{21}^{0\mathrm{R}}(t,0)^{\ast}=\frac{\mathrm{i}%
}{\hbar}\sin\left(  \frac{\omega_{0}}{2}t\right)
\end{equation}
are the isolated system propagators. The reduced density evolution is now,%
\begin{multline}
\mathbf{G}^{<}\left(  t,t\right)  =\hbar^{2}\mathbf{G}^{0\mathrm{R}}\left(
t,0\right)  \mathbf{G}^{<}\left(  0,0\right)  \mathbf{G}^{0\mathrm{A}}\left(
0,t\right)  e^{-t/\left(  2\tau_{\mathrm{SE}}\right)  }%
\label{Danielewicz_GLBE}\\
+\int_{0}^{t}\mathrm{d}t_{\mathrm{i}}\mathbf{G}^{0\mathrm{R}}\left(
t,t_{\mathrm{i}}\right)  \mathbf{\Sigma}_{\mathrm{sym.}}^{<}\left(
t_{\mathrm{i}}\right)  \mathbf{G}^{0\mathrm{A}}\left(  t_{\mathrm{i}%
},t\right)  e^{-\left(  t-t_{\mathrm{i}}\right)  /\left(  2\tau_{\mathrm{SE}%
}\right)  },
\end{multline}
which is similar to the GLBE \cite{GLBE1,GLBE2}. It is easy to see that the
introduction of negative (positive) imaginary parts in the diagonal energies
of the effective Hamiltonian produces decay (growth) rates of the elements of
the density function which, being fictitious, must be compensated by a
fictitious injection self-energy
\begin{equation}
\Sigma_{\mathrm{fic.}ij}^{<}{}\left(  t_{\mathrm{i}}\right)  =-\hbar
\operatorname{Im}\left(  \Sigma_{\mathrm{fic.}ii}^{\mathrm{R}}+\Sigma
_{\mathrm{fic.}}^{\mathrm{R}}{}_{jj}\right)  G_{ij}^{<}\left(  t_{\mathrm{i}%
},t_{\mathrm{i}}\right)  .
\end{equation}
In our case, this results in an injection that includes the compensation
effects for the symmetrized interaction%
\begin{align}
\mathbf{\Sigma}_{\mathrm{sym.}}^{<}\left(  t_{\mathrm{i}}\right)   &
=\mathbf{\Sigma}_{{}}^{<}\left(  t_{\mathrm{i}}\right)  +\mathbf{\Sigma
}_{\mathrm{fic.}}^{<}\left(  t_{\mathrm{i}}\right) \nonumber\\
&  =\left(
\begin{array}
[c]{cc}%
0 & 0\\
0 & \ \ \ \ \ 2\Gamma_{V}^{{}}\hbar\left(  \frac{\mathrm{i}}{\hbar}%
\mathrm{f}_{3}^{{}}\right)  +2\Gamma_{U}^{{}}\hbar G_{22}^{<}\left(
t_{\mathrm{i}},t_{\mathrm{i}}\right)
\end{array}
\right) \nonumber\\
&  +\left(
\begin{array}
[c]{cc}%
\Gamma_{\mathrm{SE}}\hbar G_{11}^{<}\left(  t_{\mathrm{i}},t_{\mathrm{i}%
}\right)  & 0\\
0 & \ \ \ \ \ -\Gamma_{\mathrm{SE}}\hbar G_{22}^{<}\left(  t_{\mathrm{i}%
},t_{\mathrm{i}}\right)
\end{array}
\right)  . \label{Sigma_sym}%
\end{align}
Here, the second term is proportional to the system density functions
$G_{ii}^{<}\left(  t_{\mathrm{i}},t_{\mathrm{i}}\right)  $ injecting and
extracting density on sites $1$ and $2,$ respectively, to restore the real
occupation. It is important to remark that the escape $\Gamma_{V}$ given by
$V_{23}$ in Hamiltonian (\ref{HSE}) or the process of current leads of Refs.
\cite{GLBE2,Almbladh}\ are only compensated at a constant rate by the
reservoirs. In this case, the injection self-energy is proportional to the
density function in the environment. In contrast, for voltage probes,
electron-phonon self-energies (as in Ref. \cite{GLBE2}) or our Coulombic
$\Gamma_{U}$ require an immediate charge compensation. The same is true for
the fictitious processes and this is indeed the situation of Eq.
(\ref{Sigma_sym}) where the injection self-energy is proportional to the
instantaneous system density function. Thus, the fictitious injection
self-energy compensates \emph{instantaneously} the fictitious leak and
injection of Eq. (\ref{Hsym}). This is more easily seen once Eq.
(\ref{Danielewicz_GLBE}) is integrated into a Trotter-type form, as we will
discuss in connection to Eq. (\ref{G_1step}). The symmetrization method is, in
essence, a redistribution of terms in the evolution equation
(\ref{Danielewicz_evol}) that has a simpler resolution.

We can rewrite the last expression to separate the processes that involve
density relaxation (through injection and escape processes) and pure
decoherence (through local energy fluctuations) as follows%
\begin{align}
\mathbf{\Sigma}_{\mathrm{sym.}}^{<}\left(  t_{\mathrm{i}}\right)   &
=\Sigma_{\mathrm{i}}^{<}\left(  t_{\mathrm{i}}\right)  +\Sigma_{\mathrm{m}%
}^{<}\left(  t_{\mathrm{i}}\right) \nonumber\\
&  =\mathrm{i}\Gamma_{\mathrm{SE}}\left[  2p_{V}\left(
\begin{array}
[c]{cc}%
0 & 0\\
0 & \left(  \mathrm{f}_{3}-\frac{\hbar}{\mathrm{i}}G_{22}^{<}\left(
t_{\mathrm{i}},t_{\mathrm{i}}\right)  \right)
\end{array}
\right)  \right. \nonumber\\
&  \left.  +\left(
\begin{array}
[c]{cc}%
\frac{\hbar}{\mathrm{i}}G_{11}^{<}\left(  t_{\mathrm{i}},t_{\mathrm{i}}\right)
& 0\\
0 & \frac{\hbar}{\mathrm{i}}G_{22}^{<}\left(  t_{\mathrm{i}},t_{\mathrm{i}%
}\right)
\end{array}
\right)  \right]  . \label{sigma_stroboscopio}%
\end{align}
Here
\begin{equation}
\frac{\hbar}{\mathrm{i}}G_{22}^{<}\left(  t_{\mathrm{i}},t_{\mathrm{i}%
}\right)  \equiv\frac{\hbar}{\mathrm{i}}\int G_{22}^{<}\left(  \varepsilon
,t_{\mathrm{i}}\right)  \frac{\mathrm{d}\varepsilon}{2\pi\hbar}=\mathrm{f}%
_{2}\left(  t_{\mathrm{i}}\right)
\end{equation}
and
\begin{equation}
\frac{\hbar}{\mathrm{i}}G_{11}^{<}\left(  t_{\mathrm{i}},t_{\mathrm{i}%
}\right)  =\mathrm{f}_{1}\left(  t_{\mathrm{i}}\right)  ,
\end{equation}
while, remembering that according to Eqs. (\ref{GammaSE}) and
(\ref{GammaU+GammaV}), $\Gamma_{\mathrm{SE}}=\Gamma_{U}+\Gamma_{V},$ we
define
\begin{equation}
p_{V}=\Gamma_{V}/\Gamma_{\mathrm{SE}}%
\end{equation}
as the weight of the tunneling rate relative to the total SE interaction rate.
Since the initial state has the site $2$ occupied we have that
\begin{equation}
\frac{\hbar}{\mathrm{i}}G_{ij}^{<}\left(  0,0\right)  =\delta_{i2}\delta_{2j}.
\end{equation}
Introducing Eq. (\ref{sigma_stroboscopio}) into Eq. (\ref{Danielewicz_GLBE})
and using
\begin{equation}
\frac{1}{\tau_{\mathrm{SE}}}\equiv\tfrac{2}{\hbar}\Gamma_{\mathrm{SE}},
\end{equation}
we get two coupled equations for $G_{11}^{<}$ and $G_{22}^{<}$ as follows:%
\begin{multline}
\tfrac{\hbar}{\mathrm{i}}G_{11}^{<}\left(  t,t\right)  =\left\vert \hbar
G_{12}^{0\mathrm{R}}\left(  t,0\right)  \right\vert ^{2}e^{-t/\left(
2\tau_{\mathrm{SE}}\right)  }+\label{GLBE_probability1}\\
\int\left\vert \hbar G_{12}^{0\mathrm{R}}\left(  t,t_{\mathrm{i}}\right)
\right\vert ^{2}e^{-\left(  t-t_{\mathrm{i}}\right)  /\left(  2\tau
_{\mathrm{SE}}\right)  }~2p_{V}~\frac{\mathrm{d}t_{\mathrm{i}}}{2\tau
_{\mathrm{SE}}}\left[  \mathrm{f}_{3}-\tfrac{\hbar}{\mathrm{i}}G_{22}%
^{<}\left(  t_{\mathrm{i}},t_{\mathrm{i}}\right)  \right] \\
+\int\left\vert \hbar G_{11}^{0\mathrm{R}}\left(  t,t_{\mathrm{i}}\right)
\right\vert ^{2}e^{-(t-t_{\mathrm{i}})/\left(  2\tau_{\mathrm{SE}}\right)
}\frac{\mathrm{d}t_{\mathrm{i}}}{2\tau_{\mathrm{SE}}}\left[  \tfrac{\hbar
}{\mathrm{i}}G_{11}^{<}\left(  t_{\mathrm{i}},t_{\mathrm{i}}\right)  \right]
\\
+\int\left\vert \hbar G_{12}^{0\mathrm{R}}\left(  t,t_{\mathrm{i}}\right)
\right\vert ^{2}e^{-(t-t_{\mathrm{i}})/\left(  2\tau_{\mathrm{SE}}\right)
}\frac{\mathrm{d}t_{\mathrm{i}}}{2\tau_{\mathrm{SE}}}\left[  \tfrac{\hbar
}{\mathrm{i}}G_{22}^{<}\left(  t_{\mathrm{i}},t_{\mathrm{i}}\right)  \right]
,
\end{multline}%
\begin{multline}
\tfrac{\hbar}{\mathrm{i}}G_{22}^{<}\left(  t,t\right)  =\left\vert \hbar
G_{22}^{0\mathrm{R}}\left(  t,0\right)  \right\vert ^{2}e^{-t/\left(
2\tau_{\mathrm{SE}}\right)  }+\label{GLBE_probability2}\\
\int\left\vert \hbar G_{22}^{0\mathrm{R}}\left(  t,t_{\mathrm{i}}\right)
\right\vert ^{2}e^{-\left(  t-t_{\mathrm{i}}\right)  /\left(  2\tau
_{\mathrm{SE}}\right)  }~2p_{V}~\frac{\mathrm{d}t_{\mathrm{i}}}{2\tau
_{\mathrm{SE}}}\left[  \mathrm{f}_{3}-\tfrac{\hbar}{\mathrm{i}}G_{22}%
^{<}\left(  t_{\mathrm{i}},t_{\mathrm{i}}\,\right)  \right] \\
+\int\left\vert \hbar G_{21}^{0\mathrm{R}}\left(  t,t_{\mathrm{i}}\right)
\right\vert ^{2}e^{-(t-t_{\mathrm{i}})/\left(  2\tau_{\mathrm{SE}}\right)
}\frac{\mathrm{d}t_{\mathrm{i}}}{2\tau_{\mathrm{SE}}}\left[  \tfrac{\hbar
}{\mathrm{i}}G_{11}^{<}\left(  t_{\mathrm{i}},t_{\mathrm{i}}\right)  \right]
\\
+\int\left\vert \hbar G_{22}^{0\mathrm{R}}\left(  t,t_{\mathrm{i}}\right)
\right\vert ^{2}e^{-(t-t_{\mathrm{i}})/\left(  2\tau_{\mathrm{SE}}\right)
}\frac{\mathrm{d}t_{\mathrm{i}}}{2\tau_{\mathrm{SE}}}\left[  \tfrac{\hbar
}{\mathrm{i}}G_{22}^{<}\left(  t_{\mathrm{i}},t_{\mathrm{i}}\right)  \right]
.
\end{multline}
In each equation, the first term is the probability that a particle initially
at site $2$ be found in site $1$ (or $2$) at time $t$ having survived the
interactions with the environment with a probability $e^{-t/\left(
2\tau_{\mathrm{SE}}\right)  }$. The second term describes the process of
injection/escape of particles enabled by the hopping from/towards the
reservoir, where the last of such processes occurred in the time range
($t_{\mathrm{i}},t_{\mathrm{i}}+\mathrm{d}t_{i}$) with a probability
$2p_{V}\frac{\mathrm{d}t_{\mathrm{i}}}{2\tau_{\mathrm{SE}}}.$ The injection
(escape) at site $2$ fills (empties) the site to level it up to the occupation
factor $\mathrm{f}_{3}$. The third and fourth terms take into account the last
process of measurement at time $t_{\mathrm{i}}$ due to the SE interaction with
a probability $\tfrac{\mathrm{d}t_{\mathrm{i}}}{2\tau_{\mathrm{SE}}}$. This
confirms our interpretation that in Eq. (\ref{sigma_stroboscopio}) the
dissipation processes are in $\mathbf{\Sigma}_{\mathrm{i}}^{<}\left(
t\right)  $ while $\mathbf{\Sigma}_{\mathrm{m}}^{<}\left(  t\right)  $
involves pure decoherence. It is clear that by iterating this formula, one
gets a series in the form represented in Fig. \ref{Fig_system_feynman}%
(d)\emph{.}

In summary, Eqs. (\ref{GLBE_probability1}) and (\ref{GLBE_probability2}) are
valid within the following assumptions: (a) The system is in the high
temperature limit $V_{12},U_{23},V_{23},V_{\mathrm{B}}\ll k_{\mathrm{B}}T$.
(b) The environment is assumed to have a very fast dynamic as compared with
that of the system $V_{12},U_{23},V_{23}\ll V_{\mathrm{B}}$ (fast fluctuation
regime). This is achieved through a specific model of the environment with a
very wide band where this property shows up as a flat and broad local density
of state at the \textquotedblleft surface\textquotedblright\ site $3$. These
two are the central assumptions. The consequences of them are the following
limits: (c) The self-energy is described by the self-consistent second-order
term. (d) The environment remains in equilibrium. (e) The retarded and
advanced Green's functions are calculated from a non-Hermitian effective
Hamiltonian which is independent of energy and constant in time.

The previous conditions allow us to develop a different strategy for the
solution of the spatially inhomogeneous evolution equation (GLBE): We
fictitiously symmetrize the effective Hamiltonian to impose an homogeneous
decay of the coherent dynamics. Consequently, we compensate the resulting
artificial injections and/or leaks based on a fictitious part in the injection self-energy.

\subsection{Dynamics of a swapping gate}

The solution of the coupled Eqs. (\ref{GLBE_probability1}) and
(\ref{GLBE_probability2}) involves a Laplace transform. We consider a
parameter range compatible with the spin problem where $\mathrm{f}_{3}%
\lesssim1$ while we allow the tunneling relative weight $p_{V}$ in the range
$\left[  0,1\right]  $.$\ $In a compact notation, the density function results
as follows%
\begin{equation}
\tfrac{\hbar}{\mathrm{i}}G_{11}^{<}\left(  t,t\right)  =1-a_{0}e^{-R_{0}%
t}-a_{1}\cos\left[  \left(  \omega+\mathrm{i}\eta\right)  t+\varphi
_{0}\right]  e^{-R_{1}t}. \label{G11}%
\end{equation}
Here, the decay rates $R_{0},~R_{1},$ and $\eta$, and the oscillation
frequency $\omega$ are real numbers associated with poles of the Laplace
transform. The amplitude $a_{0}$ is also real while, when $\omega=0,$ the
amplitude $a_{1}$ and the initial phase $\varphi_{0}$ acquire an imaginary
component that warrants a real density.
These observables have expressions in terms of adimensional functions of the
fundamental parameters in the model. Denoting
\begin{equation}
x=\omega_{0}\tau_{\mathrm{SE}},
\end{equation}
and remembering that
\begin{equation}
p_{V}=\Gamma_{V}/\Gamma_{\mathrm{SE}},
\end{equation}
we define
\begin{equation}
\phi\left(  p_{V},x\right)  =\frac{1}{3}\left(  x^{2}-p_{V}^{2}-\frac{1}%
{3}\left(  1-p_{V}\right)  ^{2}\right)  ,
\end{equation}
and
\begin{multline}
\chi\left(  p_{V},x\right)  =\left\{  4\left(  1-p_{V}\right)  \left(
9x^{2}-2\left(  1-p_{V}\right)  ^{2}+18p_{V}^{2}\right)  \right. \\
\left.  +12\left[  3\left(  4x^{6}-\left(  \left(  1-p_{V}\right)
^{2}+12p_{V}^{2}\right)  x^{4}\right.  \right.  \right. \\
\left.  \left.  \left.  +4p_{V}^{2}\left(  5\left(  1-p_{V}\right)
^{2}+3p_{V}^{2}\right)  x^{2}\right.  \right.  \right. \\
\left.  \left.  \left.  -4p_{V}^{2}\left(  \left(  1-p_{V}\right)  ^{2}%
-p_{V}^{2}\right)  ^{2}\right)  \right]  ^{\frac{1}{2}}\right\}  ^{\frac{1}%
{3}}.
\end{multline}
The observable \textquotedblleft frequency\textquotedblright,
\begin{equation}
\omega+\mathrm{i}\eta=\frac{\sqrt{3}}{2x}\left(  \frac{1}{6}\chi\left(
p_{V},x\right)  +6\frac{\phi\left(  p_{V},x\right)  }{\chi\left(
p_{V},x\right)  }\right)  \omega_{0}, \label{w}%
\end{equation}
is purely real or imaginary, i.e. $\omega\eta\equiv0$. Also,
\begin{equation}
R_{0}=\left(  6\frac{\phi\left(  p_{V},x\right)  }{\chi\left(  p_{V},x\right)
}-\frac{1}{6}\chi\left(  p_{V},x\right)  +p_{V}+\frac{1}{3}\left(
1-p_{V}\right)  \right)  \frac{1}{\tau_{\mathrm{SE}}},
\end{equation}%
\begin{equation}
R_{1}=\frac{3}{2}\left(  p_{V}+\frac{1}{3}\left(  1-p_{V}\right)  \right)
\frac{1}{\tau_{\mathrm{SE}}}-\frac{R_{0}}{2},
\end{equation}
and
\begin{align}
a_{0}  &  =\frac{1}{2}\frac{2\left(  \omega_{{}}^{2}-\eta_{{}}^{2}\right)
+2R_{1}^{2}-\omega_{0}^{2}}{\left(  \omega_{{}}^{2}-\eta_{{}}^{2}\right)
+\left(  R_{0}^{{}}-R_{1}^{{}}\right)  ^{2}},\\
a_{2}  &  =\dfrac{1}{2\left(  \omega+\mathrm{i}\eta\right)  }\nonumber\\
&  \times\frac{\left(  2R_{0}^{{}}R_{1}^{{}}-\omega_{0}^{2}\right)  \left(
R_{0}-R_{1}\right)  +2\left(  \omega_{{}}^{2}-\eta_{{}}^{2}\right)  R_{0}^{{}%
}}{\left(  \omega_{{}}^{2}-\eta_{{}}^{2}\right)  +\left(  R_{0}^{{}}-R_{1}%
^{{}}\right)  ^{2}},\\
a_{3}  &  =\frac{1}{2}\frac{\omega_{0}^{2}+2R_{0}^{2}-4R_{0}^{{}}R_{1}^{{}}%
}{\left(  \omega_{{}}^{2}-\eta_{{}}^{2}\right)  +\left(  R_{0}^{{}}-R_{1}^{{}%
}\right)  ^{2}},\\
a_{1}^{2}  &  =a_{2}^{2}+a_{3}^{2},\;\;\;\;\;\tan\left(  \phi_{0}\right)
=-\frac{a_{2}}{a_{3}}.
\end{align}
The oscillation frequency $\omega$ in Eq. (\ref{w}) has a critical point
$x_{\mathrm{c}}$ at a finite value of $x$ showing a quantum dynamical phase
transition for which $\omega$ and $\eta$ in Eq. (\ref{G11}) exchange their
roles as being zero and having a finite value, respectively. A full discussion
of this issue for a spin system is presented in Ref. \cite{JCP2006}. Here, the
dynamical behavior changes from a swapping phase to an overdamped phase. This
last regime can be associated with the quantum Zeno effect \cite{Misra}
where\ frequent projective measurements prevent the quantum evolution. Here,
this is a dynamical effect \cite{UsajZENO, PascazioZENO} produced by
interactions with the environment that freeze the system oscillation.

\begin{figure}
[th]
\begin{center}
\includegraphics[
trim=0.000000in 0.000000in -0.363873in 0.000000in,
height=2.5668in,
width=3.6711in
]%
{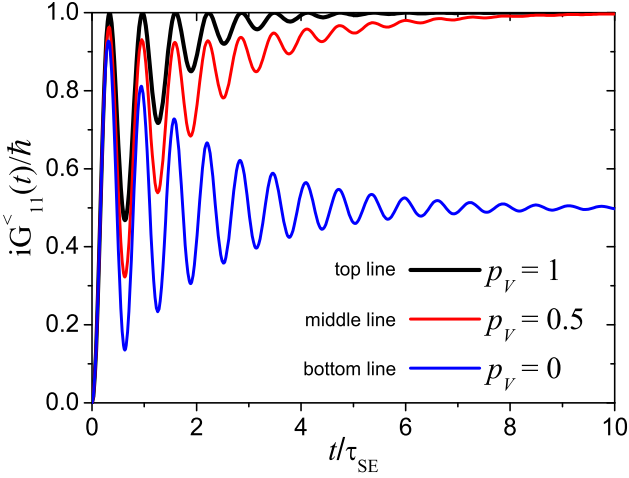}%
\caption{(Color online) Occupation probability \textrm{i}$G_{11}^{<}\left(
t\right)  /\hbar$ to find at site $1$ a particle initially at site $2.$ Each
line corresponds to different kinds, $p_{V},$ of SE interactions. The plots
correspond to $x=V_{12}\tau_{\mathrm{SE}}/\hbar=10$ belonging to the swapping
phase and \textrm{f}$_{3}=1$.}%
\label{Fig_G11}%
\end{center}
\end{figure}

Figure \ref{Fig_G11} shows typical curves of $\tfrac{\hbar}{\mathrm{i}}%
G_{11}^{<}\left(  t,t\right)  $ in the swapping phase. The different colors
correspond to different SE interactions, $p_{V}=0$, $0.5,$ and $1,$ which are
Coulomb $\left(  \Gamma_{V}=0\right)  $, isotropic $\left(  \Gamma_{V}%
=\Gamma_{U}\right)  ,$ and pure tunneling $\left(  \Gamma_{U}=0\right)  $
interactions rates. The hopping interaction does not conserve the net energy
in the system inducing a dissipation which is manifested through the
nonconservation of the number of particles in the system. This is the case of
$p_{V}\neq0$ where the final state of the system has the occupation
probability of the sites equilibrated with the bath occupation (\textrm{f}%
$_{3}$). \ In Fig. \ref{Fig_G11}, this is manifested as the asymptotic
normalized density (occupation probability) of $1.$ However, if $p_{V}=0,$
tunneling is forbidden and the system goes to an internal quasiequilibrium,
i.e., the local excitation is spread inside the system. In this case the
asymptotic occupation probability of site $1$ is $1/2.$%

\section{Stroboscopic representation of the interaction processes}

Equation. (\ref{GLBE-inhomogeneous}) has two main difficulties for a numerical
implementation: The first is the evaluation of the system nonunitary
propagators under inhomogeneous perturbations; \ the second is to keep track
of all the previous states of the system to enable the integration over past
times. We will show that the decay homogenization enables the implementation
of an efficient numerical algorithm. First of all, we identify in expression
(\ref{Danielewicz_GLBE}) that $e^{-t/\left(  2\tau_{\mathrm{SE}}\right)
}=s\left(  t\right)  $ is the system's survival probability to the environment
interruption, i.e., the probability that the system remains coherent, and
$\mathrm{d}t_{\mathrm{i}}/(2\tau_{\mathrm{SE}})=q\left(  t_{\mathrm{i}%
}\right)  \mathrm{d}t_{\mathrm{i}}$ is the \textquotedblleft
interruption\textquotedblright\ probability in a differential time around
$t_{\mathrm{i}}$. The interaction of the environment is discretized in
intervals $\tau_{\mathrm{str.}}$ where it acts instantaneously. This
stroboscopic interaction leads to%
\begin{align}
s\left(  t\right)   &  =\left(  1-p\right)  ^{n\left(  t\right)
},\label{survivep}\\
q\left(  t\right)   &  =\sum_{m=1}^{\infty}p~\delta\left(  t-m\tau
_{\mathrm{str.}}\right)  , \label{interruptionp}%
\end{align}
where
\begin{equation}
n\left(  t\right)  =\mathrm{int}\left(  t/\tau_{\mathrm{str.}}\right)  .
\end{equation}
Here, the stroboscopic interruptions may occur at the discrete times
$m\tau_{\mathrm{str.}}$ with a probability $p$. At time $t$ there were
$n\left(  t\right)  $ possible interruptions. In the joint limit
$\tau_{\mathrm{str.}}\rightarrow0$ and $p\rightarrow0$ such that
\begin{equation}
p/\tau_{\mathrm{str.}}=1/\left(  2\tau_{\mathrm{SE}}\right)  ,
\end{equation}
we recover the continuous expression (see the Appendix).

Introducing Eqs. (\ref{survivep}) and (\ref{interruptionp}) into the reduced
density expression (\ref{Danielewicz_GLBE}) we obtain%
\begin{multline}
\mathbf{G}^{<}\left(  t,t\right)  =\hbar^{2}\mathbf{G}^{0\mathrm{R}}\left(
t,0\right)  \mathbf{G}^{<}\left(  0,0\right)  \mathbf{G}^{0\mathrm{A}}\left(
0,t\right)  \left(  1-p\right)  ^{n(t)}\\
+\int_{0}^{t}\mathrm{d}t_{\mathrm{i}}\tau_{\mathrm{SE}}\sum_{m=1}^{\infty
}\delta\left(  t_{\mathrm{i}}-t_{m}\right) \\
\times\mathbf{G}^{0\mathrm{R}}\left(  t,t_{\mathrm{i}}\right)  \mathbf{\Sigma
}_{\mathrm{sym.}}^{<}\left(  t_{\mathrm{i}}\right)  \mathbf{G}^{0\mathrm{A}%
}\left(  t_{\mathrm{i}},t\right)  p\left(  1-p\right)  ^{n(t-t_{\mathrm{i}})},
\end{multline}
and rewriting we have%
\begin{multline}
\mathbf{G}^{<}\left(  t,t\right)  =\hbar^{2}\mathbf{G}^{0\mathrm{R}}\left(
t,0\right)  \mathbf{G}^{<}\left(  0,0\right)  \mathbf{G}^{0\mathrm{A}}\left(
0,t\right)  \left(  1-p\right)  ^{n}\label{GLBE_stroboscopic}\\
+\hbar^{2}\sum_{m=1}^{n}\mathbf{G}^{0\mathrm{R}}\left(  t,t_{m}\right)
\delta\mathbf{G}_{\mathrm{inj.}}^{<}\left(  t_{m},t_{m}\right)  \mathbf{G}%
^{0\mathrm{A}}\left(  t,t_{m}\right) \\
\times p\left(  1-p\right)  ^{n-m},
\end{multline}
where $n=n(t),$ $t_{m}=m\tau_{\mathrm{str.}},$ and
\begin{equation}
\delta\mathbf{G}_{\mathrm{inj.}}^{<}\left(  t,t\right)  =\frac{2\tau
_{\mathrm{SE}}}{\hbar^{2}}\mathbf{\Sigma}_{\mathrm{sym.}}^{<}\left(  t\right)
.
\end{equation}
In this picture, the evolution between interruptions is governed by the
system's\ propagators
\begin{equation}
\mathbf{G}^{0\mathrm{R}}\left(  t,0\right)  =-\frac{\mathrm{i}}{\hbar}%
\exp[-\mathrm{i}\mathbf{H}_{\mathrm{S}}t/\hbar]
\end{equation}
and
\begin{equation}
\mathbf{G}^{0\mathrm{A}}\left(  0,t\right)  =\mathbf{G}^{0\mathrm{R}}\left(
t,0\right)  ^{\dag}.
\end{equation}
The spin bath stroboscopically interrupts the system evolution producing the
decay of the coherent beam. This decay is compensated through the reinjection
of probability (or eventually of coherences) expressed in the
\textit{instantaneous interruption function} $\delta\mathbf{G}_{\mathrm{inj.}%
}^{<}\left(  t,t\right)  $ which also contains actual injection (decay) from
(to) the bath.

\begin{figure}
[th]
\begin{center}
\includegraphics[
trim=0.000000in 0.000000in -0.055141in 0.000000in,
height=3.5293in,
width=3.346in
]%
{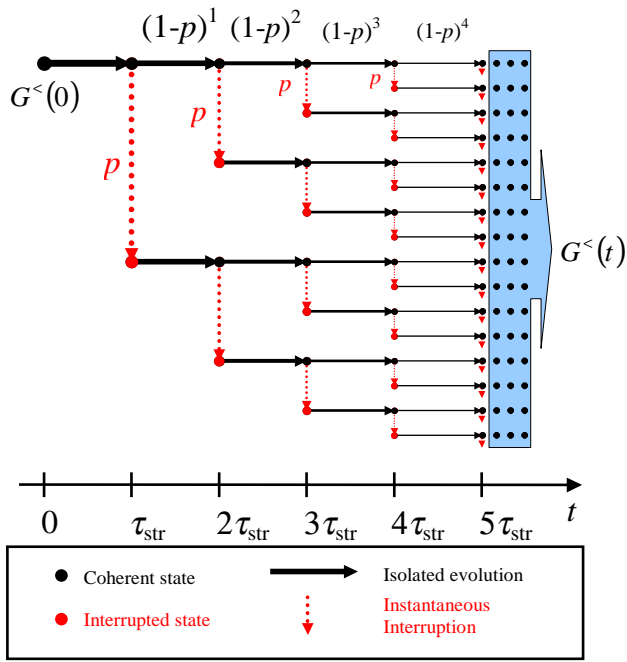}%
\caption{(Color online){} Quantum branching sequence for the stroboscopic
evolution. Gray (red) dots represent states with interrupted (incoherent)
evolution while the black dots are coherent with their predecessor. The
horizontal continuous arrows represent the isolated evolution and the vertical
dashed lines are the instantaneous interruptions. Notice the self-similar
structure.}%
\label{Fig_stroboscopy}%
\end{center}
\end{figure}

The first term in the rhs of Eq. (\ref{GLBE_stroboscopic}) is the coherent
system evolution weighted by its survival probability $\left(  1-p\right)
^{n}.$ This is the upper branch in Fig. \ref{Fig_stroboscopy}. The second term
is the incoherent evolution involving all the decoherent branches. The $m$-th
term in the sum represents the evolution that had its \textit{last}
interruption at $m\tau_{\mathrm{str.}}$ and since then survived coherently
until $n\tau_{\mathrm{str.}}$. Each of these terms is represented in Fig.
\ref{Fig_stroboscopy} by all the branches with an interrupted state (gray dot,
red online) at the hierarchy level $m$ after which they survive without
further interruptions until $n\tau_{\mathrm{str.}}$. This representation has
an immediate resemblance to that introduced by Pascazio and Namiki to justify
the dynamical Zeno effect \cite{PascazioZENO}.%
\emph{ }

As mentioned above, the solutions of Eqs. (\ref{GLBE_stroboscopic})
and\ (\ref{Danielewicz_GLBE}) are both computationally demanding since they
involve the storage of all the previous steps and reiterated summations. Thus,
taking advantage of the self-similarity of the hierarchy levels in the
interaction with the environment, we rearrange expression
(\ref{GLBE_stroboscopic}) into a form optimized for numerical computation,%
\begin{multline}
\tfrac{1}{\hbar^{2}}\mathbf{G}^{<}\left(  t_{n+1},t_{n+1}\right)
=\label{G_1step}\\
\mathbf{G}^{0\mathrm{R}}\left(  t_{n+1},t_{n}\right)  \mathbf{G}^{<}\left(
t_{n},t_{n}\right)  \mathbf{G}^{0\mathrm{A}}\left(  t_{n},t_{n+1}\right)
\left(  1-p\right) \\
+\mathbf{G}^{0\mathrm{R}}\left(  t_{n+1},t_{n}\right)  \delta\mathbf{G}%
_{\mathrm{inj.}}^{<}\left(  t_{n},t_{n}\right)  \mathbf{G}^{0\mathrm{A}%
}\left(  t_{n},t_{n+1}\right)  p.
\end{multline}
This equation provides a new computational procedure that only requires the
storage of the density function at a single previous step. Besides, it avoids
random averages required in models that include decoherence through stochastic
or kicked-like perturbations \cite{paz2003,molmer1992}. This strategy is being
implemented in our group in various cases involving quantum dynamics of many
spin systems in the presence of dissipation processes and
decoherence.\ Equation (\ref{G_1step}) manifests that the fictitious
self-energy, proportional to $G_{ii}^{<}\left(  t_{n},t_{n}\right)  $,
compensates instantaneously the fictitious leaks and injection of Eq.
(\ref{Hsym}). The conceptual consistency of the approach is illustrated by
choosing $\delta\mathbf{G}_{\mathrm{inj.}}^{<}\left(  t_{n},t_{n}\right)
\equiv\mathbf{G}^{<}\left(  t_{n},t_{n}\right)  $: one recovers a coherent
isolated evolution.

\section{Application to spin systems}

We apply this procedure to the spin system of Ref. \cite{JCP2006} providing a
first principle derivation of the phenomenological equations employed there.
We consider a system with $M=2$ spins $1/2$ coupled to a spin environment with
the following Hamiltonian $\widehat{\mathcal{H}}=\widehat{\mathcal{H}%
}_{\mathrm{S}}+\widehat{\mathcal{H}}_{\mathrm{E}}+\widehat{\mathcal{H}%
}_{\mathrm{SE}}$, where the system Hamiltonian $\widehat{\mathcal{H}%
}_{\mathrm{S}}$ is
\begin{align}
\widehat{\mathcal{H}}_{\mathrm{S}}  &  =\hbar\omega_{\mathrm{L}}^{{}}\left(
\hat{I}_{1}^{z}+\hat{I}_{2}^{z}\right)  +\tfrac{1}{2}b_{12}^{{}}\left(
\hat{I}_{1}^{+}\hat{I}_{2}^{-}+\hat{I}_{1}^{-}\hat{I}_{2}^{+}\right)
\nonumber\\
&  =\hbar\omega_{\mathrm{L}}^{{}}\left(  \hat{I}_{1}^{z}+\hat{I}_{2}%
^{z}\right)  +b_{12}^{{}}\left(  \hat{I}_{1}^{x}\hat{I}_{2}^{x}+\hat{I}%
_{1}^{y}\hat{I}_{2}^{y}\right)  .
\end{align}
Here, the first term is the Zeeman energy and the second term gives a
flip-flop or $XY$ spin-spin interaction\emph{. }The environment Hamiltonian is
described by%
\begin{equation}
\widehat{\mathcal{H}}_{\mathrm{E}}=\sum_{i\geq3}\hbar\omega_{\mathrm{L}}^{{}%
}\hat{I}_{i}^{z}+%
{\textstyle\sum\limits_{\genfrac{}{}{0pt}{1}{i\geq3}{j>i}}}
\tfrac{1}{2}b_{ij}^{{}}\left(  \hat{I}_{i}^{+}\hat{I}_{j}^{-}+\hat{I}_{i}%
^{-}\hat{I}_{j}^{+}\right)  ,
\end{equation}
and for the SE interaction we have%
\begin{equation}
\widehat{\mathcal{H}}_{\mathrm{SE}}=a_{23}^{{}}\hat{I}_{2}^{z}\hat{I}_{3}%
^{z}+\tfrac{1}{2}b_{23}^{{}}\left(  \hat{I}_{2}^{+}\hat{I}_{3}^{-}+\hat{I}%
_{2}^{-}\hat{I}_{3}^{+}\right)  ,
\end{equation}
where this spin-spin interaction is Ising if $b_{23}/a_{23}=0,$ and $XY$,
isotropic (Heisenberg), or the truncated dipolar (secular) if $a_{23}%
/b_{23}=0,1,-2,$ respectively.

We map the spin system into a fermionic system using the Jordan-Wigner
transformation (JWT) \cite{spin-fermion},
\begin{equation}
\hat{I}_{i}^{+}=\hat{c}_{i}^{+}\exp\left\{  \mathrm{i}\pi\sum_{j=1}^{i-1}%
\hat{c}_{j}^{+}\hat{c}_{j}^{{}}\right\}  .
\end{equation}
The previous Hamiltonians become%
\begin{align}
\widehat{\mathcal{H}}_{\mathrm{S}}  &  =\hbar\omega_{\mathrm{L}}^{{}}\left(
\hat{c}_{1}^{+}\hat{c}_{1}^{{}}+\hat{c}_{2}^{+}\hat{c}_{2}^{{}}-\mathbf{1}%
\right)  +\tfrac{1}{2}b_{12}\left(  \hat{c}_{1}^{+}\hat{c}_{2}^{{}}+\hat
{c}_{2}^{+}\hat{c}_{1}^{{}}\right)  ,\label{HS_spins}\\
\widehat{\mathcal{H}}_{\mathrm{E}}  &  =\sum_{i\geq3}\hbar\omega_{\mathrm{L}%
}^{{}}\left(  \hat{c}_{i}^{+}\hat{c}_{i}^{{}}-\tfrac{1}{2}\mathbf{1}\right)  +%
{\textstyle\sum\limits_{\genfrac{}{}{0pt}{1}{i\geq3}{j>i}}}
\tfrac{1}{2}b_{ij}^{{}}\left(  \hat{c}_{i}^{+}\hat{c}_{j}^{{}}+\hat{c}_{j}%
^{+}\hat{c}_{i}^{{}}\right)  ,\label{HE_spins}\\
\widehat{\mathcal{H}}_{\mathrm{SE}}  &  =a_{23}^{{}}\left(  \hat{c}_{2}%
^{+}\hat{c}_{2}^{{}}-\tfrac{\mathbf{1}}{2}\right)  \left(  \hat{c}_{3}^{+}%
\hat{c}_{3}^{{}}-\tfrac{\mathbf{1}}{2}\right)  +\tfrac{1}{2}b_{23}^{{}}\left(
\hat{c}_{2}^{+}\hat{c}_{3}^{{}}+\hat{c}_{3}^{+}\hat{c}_{2}^{{}}\right)  .
\label{HSE_spins}%
\end{align}
Here, the system interacts with the environment through site $3$ (the surface
site of the bath). In the last Hamiltonians, the terms proportional to the
identity do not contribute to the dynamics because they only change the total
energy by a constant number. This Hamiltonian describes a standard cross
polarization experiment (swapping gate) in NMR \cite{MKBE74,JCP2006}. In this
experiment, site $1$ is a $^{13}$C and site $2$ is a $^{1}$H while the
environment is a $^{1}$H spin bath. The typical experimental Hartmann-Hahn
condition \cite{MKBE74,JCP2006} equals the values of the \ effective energies
at the $^{13}$C and the $^{1}$H sites to optimize the polarization transfer.
The SE interaction has terms linear in the number operators $\hat{c}_{2}%
^{+}\hat{c}_{2}^{{}}$ and $\hat{c}_{3}^{+}\hat{c}_{3}^{{}},$ that only change
the energy of sites $2$ and $3,$ respectively. Thus, the Hartmann-Hahn
implementation, compensates the change of energy produced by the environment
through these linear terms. Finally, we have Hamiltonians equivalent to those
in Eqs. (\ref{Hs}-\ref{HSE}) where the site energies are equal, and
$V_{12}=-\frac{b_{12}}{2},~V_{ij}=-\tfrac{b_{ij}}{2}$, $U_{23}%
^{\mathrm{(dir.)}}=a_{23}^{{}},$ and $U_{23}^{\mathrm{(exch.)}}=0.$

The spin dynamics of the system is described by the spin correlation function
\cite{spin-projection,CPL2005} as follows:%
\begin{equation}
P_{i2}(t)=\frac{\left\langle \Psi_{\mathrm{eq.}}\right\vert \hat{I}_{i}%
^{z}(t)\hat{I}_{2}^{z}(0)\left\vert \Psi_{\mathrm{eq.}}\right\rangle
}{\left\langle \Psi_{\mathrm{eq.}}\right\vert \hat{I}_{2}^{z}(0)\hat{I}%
_{2}^{z}(0)\left\vert \Psi_{\mathrm{eq.}}\right\rangle },
\end{equation}
which gives the local polarization at time $t$ on the $i$-th spin with an
initial local excitation on the $2$-nd spin at time $t=0.$ Here, $\left\vert
\Psi_{\mathrm{eq.}}\right\rangle $ is the thermodynamical many-body
equilibrium state and
\begin{equation}
\hat{I}_{i}^{z}(t)=e^{\mathrm{i}\widehat{\mathcal{H}}t/\hbar}\hat{I}_{i}%
^{z}e^{-\mathrm{i}\widehat{\mathcal{H}}t/\hbar}%
\end{equation}
are the spin operators in the Heisenberg representation. After the JWT, the
initial local excitation on site $2$ is described by the nonequilibrium state
\begin{equation}
\left\vert \Psi_{\mathrm{n.e.}}\right\rangle =\hat{c}_{2}^{+}\left\vert
\Psi_{\mathrm{eq.}}\right\rangle .
\end{equation}
In the experimental high temperature regime, $k_{\mathrm{B}}T$ much\ larger
than any energy scale of the system, the spin correlation function becomes%
\begin{equation}
P_{i2}(t)=\tfrac{2\hbar}{\mathrm{i}}G_{ii}^{<\,}(t,t)-1. \label{Pol_spin}%
\end{equation}
Notice that $G_{ii}^{<\,}(t,t)$ implicitly depends on the initial local
excitation at site $2$. Here, $G_{ii}^{<\,}(t,t)$ is the reduced density
function of sites $1$ and $2$ and can be split into the contributions
$G_{ii}^{<\,_{(N)}}(t_{2},t_{1})$ from each subspace with $N$ particles (or
equivalently $N$ spins up) in the following way \cite{CPL2005}:%
\begin{equation}
G_{ii}^{<\,}(t,t)=\sum_{N=1}^{M}\dfrac{\left(
\genfrac{}{}{0pt}{1}{M-1}{N-1}%
\right)  }{2^{M-1}}G_{ii}^{<\,_{(N)}}(t,t),
\end{equation}
and analogous for the hole density function. The initial condition in this
picture is described by%
\begin{equation}
G_{ij}^{<_{(N)}}(0,0)=\tfrac{\mathrm{i}}{\hbar}\left(  \tfrac{N-1}{M-1}%
\delta_{ij}+\tfrac{M-N}{M-1}\delta_{i2}\delta_{2j}\right)  ,
\label{Initial_condition_spins}%
\end{equation}
where the first term is the equilibrium density (identical occupation for all
the sites) and the second term is the nonequilibrium contribution where only
site $2$ is excited. Thus, we have an expression such as
(\ref{Danielewicz_evol}) for each $N$-th subspace (see Ref. \cite{CPL2005}).
For this two-spin system, as we showed in \cite{CPL2005}, the $-1$ term of Eq.
(\ref{Pol_spin}) is canceled out by the background evolution, i.e., the
evolution of the first term of Eq. (\ref{Initial_condition_spins}) plus the
evolution of the second term of Eq. (\ref{Danielewicz_evol}) for the $N=2$
subspace. As a consequence, the observable dynamics only depends on the
initial local excitation at site $2,$%
\begin{equation}
G_{ij}^{<_{\left(  1\right)  }}(0,0)=\tfrac{\mathrm{i}}{\hbar}\delta
_{i2}\delta_{2j},
\end{equation}
and evolves in the first particle subspace,
\begin{equation}
P_{i2}(t)=\tfrac{\hbar}{\mathrm{i}}G_{ii}^{<\,_{(1)}}(t,t).
\end{equation}
Finally, the solution of the polarization $P_{12}(t)$ is the same as that
obtained in Eq. (\ref{G11}).

By substituting in the present microscopic model $\Gamma_{\mathrm{XY}%
}\leftrightarrow\Gamma_{V}$ and $\Gamma_{\mathrm{ZZ}}\leftrightarrow\Gamma
_{U}$, we obtain the same dynamics as that found in Ref. \cite{JCP2006} for a
phenomenological spin model. There, we showed that such a solution presents a
quantum dynamical phase transition in fair agreement with the phenomenon
observed experimentally \cite{JCP98}.

\section{Conclusions}

We have shown a method that involves the transformation of the density
function expressed in the Danielewicz integral form into a generalized
Landauer-B\"{u}ttiker equation. This was possible by resorting to Wigner
energy-time variables to perform the fast fluctuation approximation for the
environment which leads to interactions local in time. Further on, we
effectively symmetrized the system-environment interactions transforming them
into a spatially homogeneous process. This has a uniform system-environment
interaction rate leading to a simple non-Hermitian propagator. The original
multiexponential decay processes are recovered by an injection density
function. Moreover, through discretization of the GLBE, we built a
stroboscopic process which is the basis for an optimal numerical algorithm
where the quantum dynamics is calculated in discrete time steps. Finally, we
applied these techniques to a spin system giving a microscopic derivation that
justifies the stroboscopic model used in Ref. \cite{JCP2006} to explain the
experimentally observed quantum dynamical phase transition.

\bigskip

\appendix

\section{Recovering the continuous process}

In order to recover the continuous expression (\ref{Danielewicz_GLBE}) from
the stroboscopic one (\ref{GLBE_stroboscopic}) we notice that if $n\left(
t\right)  =n,$ we can write Eq. (\ref{survivep}) as%
\begin{equation}
s\left(  t\right)  =\left(  1-p\right)  ^{\frac{\left(  n\tau_{\mathrm{str.}%
}\right)  }{\tau_{\mathrm{str.}}}}=\left(  1-\frac{\tau_{\mathrm{str.}}}%
{2\tau_{\mathrm{SE}}}\right)  ^{\left(  n\tau_{\mathrm{str.}}\right)
/\tau_{\mathrm{str.}}}.
\end{equation}
If $t=n\tau_{\mathrm{str.}}$ then%
\begin{equation}
s\left(  t\right)  =\left(  1-\frac{\tau_{\mathrm{str.}}}{2\tau_{\mathrm{SE}}%
}\right)  ^{t/\tau_{\mathrm{str.}}}.
\end{equation}
By taking the limit $\tau_{\mathrm{str.}}\rightarrow0$ the variable $t$
becomes continuous yielding%
\begin{align}
s\left(  t\right)   &  =\lim_{\tau_{\mathrm{str.}}\rightarrow0}\left(
1-\frac{\tau_{\mathrm{str.}}}{2\tau_{\mathrm{SE}}}\right)  ^{t/\tau
_{\mathrm{str.}}}\nonumber\\
&  =\exp\left[  -t/\left(  2\tau_{\mathrm{SE}}\right)  \right]  ,
\end{align}
recovering the continuous expression for $s\left(  t\right)  .$

By substituting $p=\tau_{\mathrm{str.}}/(2\tau_{\mathrm{SE}})$ in Eq.
(\ref{interruptionp}) we have%
\begin{equation}
q(t)=\frac{1}{2\tau_{\mathrm{SE}}}\sum_{m=1}^{\infty}\tau_{\mathrm{str.}%
}\delta(t-m\tau_{\mathrm{str.}}).
\end{equation}
In the limit $\tau_{\mathrm{str.}}\rightarrow0,$ $t_{m}=m\tau_{\mathrm{str.}}$
becomes a continuous variable and we can convert the sum into an integral,
leading to
\begin{equation}
q(t)=\frac{1}{2\tau_{\mathrm{SE}}}\int_{0}^{\infty}\tau_{\mathrm{str.}}%
\delta(t-t_{m})\frac{\mathrm{d}t_{m}}{\tau_{\mathrm{str.}}}=\frac{1}%
{2\tau_{\mathrm{SE}}}.
\end{equation}
The continuous expression of the GLBE (\ref{Danielewicz_GLBE}) is then obtained.


\begin{thebibliography}{99}
\bibitem {matrix-representation}The characters in bold font are matrix
representations of the respective operator.
\end{thebibliography}
\end{document}